\documentclass{article}
\usepackage{import}
\usepackage{tikz}
\usepackage{pgf-pie}
\usepackage{tcolorbox}
\usepackage{subcaption}
\usepackage{enumitem}
\usepackage{xcolor}
\usepackage{soul,color}
\usepackage{fnpct}
\usepackage{adjustbox}
\usepackage{graphicx}
\usepackage{xcolor}
\usepackage[multiple]{footmisc}
\usepackage{xurl}

\title{Test-takers have a say: understanding the implications of the use of AI in language tests}

\author{Dawen Zhang$^{1,2}$, Thong Hoang$^{1}$, Shidong Pan$^{1,2}$,\\ Yongquan Hu$^{3}$, Zhenchang Xing$^{1,2}$, Mark Staples$^{1}$,\\ Xiwei Xu$^{1}$, Qinghua Lu$^{1}$, Aaron Quigley$^{1}$
\\
\\
\textit{CSIRO's Data61$^1$}
\\
\textit{Australian National University$^2$}
\\
\textit{University of New South Wales$^3$}
\\
\\
\textit{David.Zhang@data61.csiro.au}
}
\date{}

\begin{document}

\maketitle

\begin{abstract}
Language tests measure a person's ability to use a language in terms of listening, speaking, reading, or writing. Such tests play an integral role in academic, professional, and immigration domains, with entities such as educational institutions, professional accreditation bodies, and governments using them to assess candidate language proficiency. Recent advances in Artificial Intelligence (AI) and the discipline of Natural Language Processing have prompted language test providers to explore AI's potential applicability within language testing, leading to transformative activity patterns surrounding language instruction and learning. However, with concerns over AI's trustworthiness, it is imperative to understand the implications of integrating AI into language testing. This knowledge will enable stakeholders to make well-informed decisions, thus safeguarding community well-being and testing integrity. To understand the concerns and effects of AI usage in language tests, we conducted interviews and surveys with English test-takers. To the best of our knowledge, this is the first empirical study aimed at identifying the implications of AI adoption in language tests from a test-taker perspective. Our study reveals test-taker perceptions and behavioral patterns. Specifically, we identify that AI integration may enhance perceptions of fairness, consistency, and availability. Conversely, it might incite mistrust regarding reliability and interactivity aspects, subsequently influencing the behaviors and well-being of test-takers. These insights provide a better understanding of potential societal implications and assist stakeholders in making informed decisions concerning AI usage in language testing.
\end{abstract}

\maketitle

\section{Introduction}
\label{sec:intro}

Language tests (LTs) have played an important, and at times contentious part in society, serving multiple roles across sectors, such as education~\cite{coley1999english}, immigration~\cite{blackledge2009country}, or citizenship~\cite{hogan2009discourses}. For example, many Australian universities necessitate that non-native English speakers attain a certain score in an accredited English language test to satisfy admission criteria~\cite{coley1999english}. Similarly, permanent residency in several countries including the United Kingdom, Canada, and Australia typically hinges on the completion of an English language test~\cite{roever2006language}. In addition, the United States (U.S.) government specifies that ``\textit{applicants must demonstrate a basic understanding of English, including an ability to read, write, and speak the language}'' to obtain U.S. citizenship~\cite{hogan2009discourses}.

LTs are ``\textit{the practice and study of evaluating the proficiency of an individual in using a particular language effectively}''~\cite{sumintono2019english}. Specifically, these tests aim to assess the language proficiency of candidates at specific levels. Educational institutions, professional accreditation bodies, and governments employ LTs to evaluate whether nominees have sufficient language skills to study abroad, engage in professional work, immigrate, or pursue naturalization.
Consequently, LTs could directly influence the areas of immigration, education, and employment, thereby affecting the well-being of individuals and, by extension, entire communities~\cite{roever2006language}. For example, LTs have been employed as ``gatekeepers'' to selectively deny entry to people at borders in the past.

Artificial Intelligence (AI) simulates human intelligence by utilizing computer systems and has become increasingly vital in contemporary society~\cite{kok2009artificial, simmons1988artificial}. AI is applied in sophisticated search engines~\cite{brin2012reprint, jansen2006we}, recommendation systems~\cite{gomez2015netflix, smith2017two}, natural language understanding~\cite{chowdhary2020natural, fitzgerald2022alexa}, and autonomous vehicles~\cite{cerf2018comprehensive, rosenband2017inside}. In language learning, developers incorporate AI models into their software applications to enhance pronunciation and vocabulary skills~\cite{samad2020elsa}, review spelling and grammar mistakes~\cite{ghufron2018role}, or assist in crafting articles for clients~\cite{godwin2022partnering}.
However with LTs, the process of human scoring can be laborious and time-consuming~\cite{weigle2013english, evanini2017approaches}, and it often presents the challenge of eliminating human bias~\cite{ferman2021discriminating, ferman2022assessing, protivinsky2018gender}. Therefore, an automated scoring system is a sought-after solution to reduce costs and minimize human error. AI has demonstrated its utility across various real-world applications, leading language test organizers to increasingly adopt AI algorithms in constructing their automated scoring systems.
For example, 
Pearson Language Tests,\footnote{\url{https://www.pearsonpte.com/}} which validate the proficiency of non-native English speakers, have implemented an automated scoring system to grade test-takers. The PTE test provider asserts that its automated scoring system, built upon a vast array of real responses from test-takers, is precise, consistent, and unbiased.\footnote{\url{https://www.pearsonpte.com/scoring/automated-scoring}}

While researchers often highlight the efficiency of AI applications, they may overlook other crucial aspects, such as AI trustworthiness and AI ethics. \textit{AI trustworthiness} involves assessing the safety and reliability of AI models based on pre-defined criteria, i.e., fairness, explainability, accountability, reliability, and acceptance~\cite{kaur2022trustworthy}. This assessment process is vital and can significantly influence the future adoption of AI applications. Over the past decade, AI trustworthiness has received considerable scholarly attention~\cite{kaur2022trustworthy, phillips2020four, ashoori2019ai, li2023trustworthy}. For example, Ashoori and Weisz~\cite{ashoori2019ai} evaluated various factors, i.e., decision stakes, decision authority, and model trainers, that could influence trust in AI. Recently, Li et al.~\cite{li2023trustworthy} argued that enhancing the trustworthiness of AI models necessitates concerted efforts at multiple stages of an AI product's lifecycle, including data preparation and algorithm design.
On the other hand, \textit{AI ethics} focuses on examining ethical challenges such as privacy issues and data bias during the design and development of AI models. Given that AI applications are now impacting diverse aspects of our lives, including healthcare~\cite{yu2018artificial}, transportation~\cite{sadek2007artificial}, and business~\cite{akerkar2019artificial}, the exploration of AI's ethical implications is a pressing need. The literature on AI ethics is substantial~\cite{jobin2019global, coeckelbergh2020ai, hagendorff2020ethics, kazim2021high}. For instance, Jobin et al.~\cite{jobin2019global} outlined a set of principles and guidelines for AI systems' ethics. Following this work, Kazim and Koshiyama~\cite{kaur2022trustworthy} proposed three research directions for AI ethics—principles, processes, and ethical consciousness—as critical focal points to address ethical issues in AI systems.

Several studies have examined the fairness of automated scoring systems, employing AI models in Language Tests (LTs)~\cite{hamid2019test, loukina2019many}. However, to our knowledge, no prior work has thoroughly explored the trustworthiness, transparency, consistency, or explainability of these systems from the test-takers' perspective. Neglecting these factors could influence the employment of AI in LTs and the derived benefits for test-takers. In this paper, we undertake an empirical analysis to assess the implications of AI usage in English LTs. Specifically, we engage with test-takers of globally recognized English exams, including the TOEFL, International English Language Testing System\footnote{\url{https://www.ielts.org/}} (IELTS), PTE, and Duolingo English Test\footnote{\url{https://englishtest.duolingo.com/}} (DET). The insights garnered will assist researchers and developers in better understanding the effective application of AI models in their evaluation systems and uncover the broader implications of current AI usage for all LT stakeholders. Our investigation seeks to answer the following research questions:

\begin{itemize}
    \item What concerns do test-takers have regarding various types of language tests?
    \item What is the impact of using AI in language tests on test-takers?
\end{itemize}

\noindent Our paper makes the following key contributions:  
\begin{itemize}
    \item We conduct an empirical study to uncover the implications of using AI in language tests, with a specific focus on the perspective of test-takers. Our interviews and online surveys offer concrete empirical evidence to underpin our findings.
    \item We outline six distinct categories of concerns and two types of impacts experienced by language test-takers.
    \item We explain the potential consequences for stakeholders involved in language tests. Furthermore, we offer guidance on incorporating AI techniques into language tests in a fair, effective, and seamless manner. 
\end{itemize}

\section{Background and Related Work}
\label{sec:background}
This section provides an overview of the history of English language tests, delves into the concept of automated scoring, and reviews the existing literature pertaining to the implications of AI implementation in these language tests.

\subsection{Language Tests}

The first formal language test was initiated by the Michigan Language Assessment in 1941 with the objective of evaluating the English proficiency of foreign students at Michigan and other universities in the United States (U.S.)~\cite{song2005language}. Subsequently, in 1964, the Test of English as a Foreign Language (TOEFL) was introduced by the Educational Testing Service (ETS). This test was designed to assess a wide range of English skills, including reading, listening, speaking, and writing, for students seeking admission to U.S. and Canadian universities~\cite{alderson2009test}. Since its inception, TOEFL has gained global recognition and evolved into an internet-based format, known as TOEFL iBT.\footnote{\url{https://www.ets.org/toefl/test-takers/ibt.html}}

With the increasing demand for language proficiency assessment, numerous language test organizations have emerged to cater to this need. For example, the International English Language Testing System (IELTS), introduced in 1989 by the British Council, is globally recognized as a reliable English language proficiency test~\cite{davidson1998book}. Moreover, IELTS is accepted by most English-speaking academic institutions and numerous international professional organizations.\footnote{\url{https://www.ieltsasia.org/hk/en/study-in-us/required-score}} In some English-speaking countries, such as the United Kingdom, Australia, and New Zealand, the IELTS certificate is also employed for visa applications.\footnote{\url{https://leapscholar.com/blog/ielts-countries-list-accepting-exam-academic/}} Differing from the TOEFL or IELTS, which offer both paper-based and computer-based formats, the Pearson Language Tests (PTE) are exclusively computer-based.\footnote{\url{https://www.pearsonpte.com/pte-academic}} In addition, the PTE employs an automated scoring system to evaluate candidate performance. In 2016, the education company Duolingo launched its own language proficiency assessment, namely the Duolingo English Test (DET)~\cite{wagner2015duolingo}. DET offers a convenient remote testing experience, allowing candidates to complete their English assessment at any time via their computers, eliminating the need for travel to physical test centers. Given the laborious and time-intensive nature of manual language test evaluations, the development of automated scoring systems has been prioritized to reduce costs and minimize human error. Several language test organizations, including ETS, Pearson, and Duolingo, have integrated AI models into their scoring systems. This integration aims to enhance the accuracy, reliability, and efficiency of evaluating test-takers' language proficiency.

\subsection{Automated Scoring}

The exploration of automated systems in language test scoring has been an area of interest for many decades, particularly in the field of Automated Essay Scoring (AES). The initial spark for AES research was ignited by Project Essay Grade, which aimed at evaluating the quality of written essays~\cite{page1966imminence}. The subsequent attention from the Educational Testing Service (ETS) fueled the intensive study of language test scoring. Specifically, ETS implemented an AES system ('e-rater') for the evaluation of Graduate Record Examinations essays, offering early insights into the limitations and effectiveness of AES~\cite{powers2002stumping}. Rudner et al.~\cite{rudner2002automated} introduced a two-score-point AES system based on Bayes' theorem, while Foltz et al.~\cite{foltz1999intelligent} proposed the Intelligent Essay Assessor, which employed Latent Semantic Analysis to evaluate essay quality through language similarities. The advent of deep learning techniques resulted in the integration of deep neural networks into AES. A noteworthy example is the method proposed by Dong et al.~\cite{dong2017attention}, which combined an attention mechanism with a recurrent convolutional neural network, demonstrating superior performance compared to traditional AES methods. Recently, Ramesh et al. offered~\cite{ramesh2022automated} a systematic review of AES research, summarizing available datasets, features, metrics, and techniques used in the field.

Due to the success of deep neural networks, the development of Automatic Speech Recognition techniques largely accelerated the automated scoring in Speaking tests~\cite{xi2010automated}. ETS has been at the forefront of this innovation, actively creating automated scoring systems for various types of tests and incorporating diverse features, including fluency, grammar, content, and structure~\cite{evanini2015automated, evanini2017approaches, zechner2015automated}. Duolingo has also developed automated scoring systems, applying an adaptive testing approach~\cite{duolingo-paper}. Their system generates scores based on both the performance of the test-taker and the difficulty level of the question. Duolingo also designs their speaking tests in an elicited speech format~\cite{vinther2002elicited}.

\subsection{Implications of the Use of AI in Language Tests}

While numerous studies have delved into various aspects of AI applications in Language Tests (LTs), no existing research has comprehensively examined the implications such as consistency, transparency, or explainability—of AI use from the perspective of the test-takers. We highlight a few notable works that are relevant to our study. Xi et al.~\cite{xi2016chinese} conducted an empirical investigation into the perception of Chinese users towards the automated scoring system in the ETS TOEFL Practice Online (TPO) service. Hamid et al.~\cite{hamid2019test}, leveraging responses from 430 IELTS test-takers, underscored that a significant proportion of test-takers felt their language proficiency was not accurately reflected in their test scores, thereby raising concerns about the fairness of language evaluation systems. Loutina et al.~\cite{loukina2019many} assessed the fairness of automated scoring systems by considering various dimensions of test-takers, such as different groups and languages. However, these previous studies only focused on a few narrow angles, and have overlooked the broader implications of AI use in LTs, which could influence AI's adoption in LTs and limit the potential benefits for test-takers. In our paper, we aim to unravel these implications, providing valuable insights for developers to more effectively integrate AI models into their automated scoring systems.

\section{Methodology}
\label{sec:method}

In this section, we present an overview of our analytical framework, designed to assess the implications of AI use in English language tests (LTs) from the test-takers' perspective. Given that the majority of questions in the Listening and Reading sections of LTs are multiple-choice and hence easily marked, \textbf{our study focuses on the Speaking and Writing sections of these tests.}

\subsection{Framework Overview}
\label{sec:method:overview}

In this study, we categorize language tests (LTs) into two types: Human-based and AI-based. \textit{Human-based} language tests (HLTs) rely on human evaluators to score the performance of test-takers, whereas \textit{AI-based} language tests (ALTs) use automated systems, eliminating the need for human intervention during the evaluation process. Our research focuses on four major English LTs: TOEFL, IELTS, PTE, and DET. The basic information of these tests is listed in Table \ref{table:basicInfo}. By our classification, TOEFL and IELTS fall under HLTs, while PTE and DET are classified as ALTs. Note that although TOEFL incorporates AI in its scoring process, due to the presence of at least one human grader involved in every question and the primary decision-making role humans play in the scoring phase, we classify TOEFL as a Human-based language test.

\begin{table}[t!]
\centering
\adjustbox{width=1.35\textwidth,center}{
\begin{tabular}{c c c c c c} 
\hline
Name of LT & Marking mode & Official AI-assisted practice tool & Exam cost & Result release time & Remarking\\
\hline
IELTS & Human & None & \$ 255 (USA) & 3-5 days & Yes (\$ 120) \\
TOEFL & Hybrid & TOEFL Practice Online & \$ 255 (USA) & 4-8 days & Yes (\$ 80-160) \\
PTE & AI & PTE Practice Tests & \$ 235 (USA)
& 2 days & Yes (\$ 120-270) \\
DET & AI & Duolingo Practice Tests & \$ 59 (USA) & 2 days & No \\
 \hline
\end{tabular}
}
\caption{Basic information of four language tests in the study.}
\label{table:basicInfo}
\end{table}

\begin{figure}[t!]
  \centering
\adjustbox{center}{
  \includegraphics[width=1.35\textwidth]{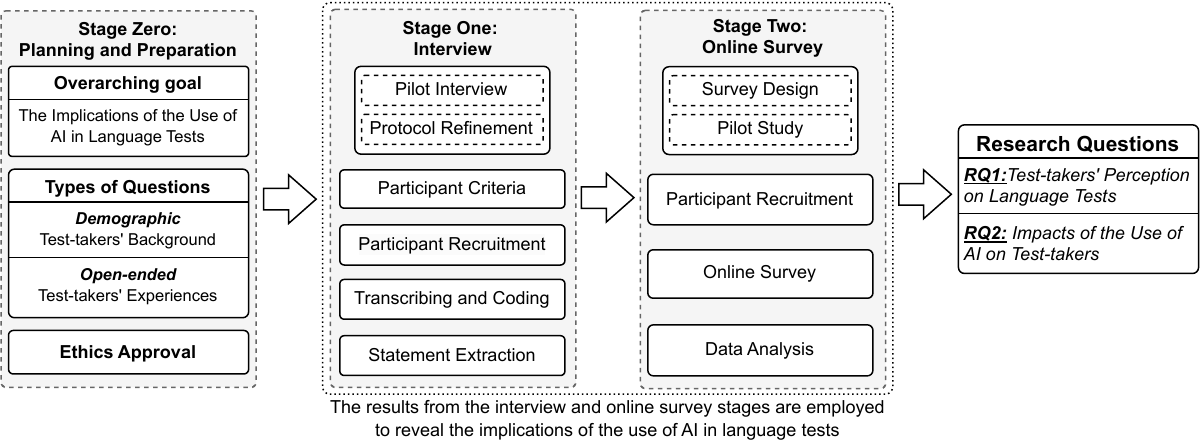}
  }
  \caption{The overview framework of our empirical study for revealing the implications of the use of AI in language tests.}
  \label{fig:methodology}
\end{figure}

Figure~\ref{fig:methodology} presents the comprehensive framework of our empirical study. Primarily, we conduct interviews with language test-takers and subsequently construct an online survey to explore their perceptions regarding the use of AI in language testing. Our framework encompasses three stages, which are detailed below:
\begin{enumerate}[start=0, leftmargin=*]
    \item \textit{Planning and Preparation.} In this initial stage, our goal is to devise a comprehensive list of questions designed to glean insights into the test-takers' backgrounds and experiences with language tests.
    \item \textit{Interview.} This stage is characterized by the conduction of interviews with test-takers. The objective here is to capture detailed, qualitative feedback on their perspectives toward LTs.
    \item \textit{Online Survey.} In the final stage, we create and distribute an online survey to a broader range of test-takers. This stage aims to capture wide-scale data on test-taker experiences and sentiments toward LTs.
\end{enumerate}
We present the details of each stage in the following subsections. 

\subsection{Stage Zero: Planning and Preparation}
\label{sec:method:planning}

In this stage, we formulate two categories of questions: demographic and open-ended. The demographic questions are intended to gather information about the test-takers' educational background and level of English proficiency. The open-ended questions are divided into two sections: general experience and AI experience. The general experience questions are designed to delve into the test-takers' previous encounters with English tests, asking which language tests they have taken and for what reasons. The AI experience questions aim to understand the test-takers' views on the use of AI in language tests, particularly concerning aspects, such as fairness, reliability, and consistency. We have secured ethics approval for this empirical study, ensuring that the test-takers retain the right to access their data and understand the process of our investigation.

\subsection{Stage One: Interview}
\label{sec:method:interview}

We introduce the interview stage of our study, designed to understand insights from test-takers' perspectives, such as consistency, transparency, and explainability, regarding English language tests (LTs). In particular, the interview stage includes the following steps. 

\noindent \textbf{Pilot Interview \& Protocol Refinement.}
Initially, we conducted pilot interviews with three individuals who had taken English LTs. We refined the interview protocol based on their feedback before launching the formal interviews.

\noindent \textbf{Participant Criteria.} 
We required our participants to have experience with at least one English language test, such as TOEFL, IELTS, PTE, or DET. In addition, participants needed to have functional English proficiency to comprehend the interview questions.

\noindent \textbf{Participant Recruitment.}
We employed social media platforms, including Twitter, Facebook, LinkedIn, and WeChat, to recruit participants. Additionally, we invited students from language schools and language test centers to join our empirical study. In the end, we have 16 participants from four different English LTs: TOEFL, IELTS, PTE, and DET. Notably, 13 participants (81.25\%) had experience with at least two different English LTs, such as IELTS and PTE. We identified the participants as I1 to I16. Table~\ref{table:interviewees} provides details about our participants.

\noindent \textbf{Transcribing and Coding.} 
We began by recording the interviews. The first author transcribed the audio, while the second author reviewed the transcripts for any English errors, i.e., mishearing, misspelling, grammar mistakes, or incorrect punctuation. Both authors then performed thematic coding on the three pilot interview transcripts for qualitative analysis.

\noindent \textbf{Statement Extraction.} 
We manually extracted statements from the transcripts, eventually deriving 25 consolidated statements representing five different stages of English LTs: test preparation, test administration, testing, test scoring, and test results. We categorized these statements into various themes, such as fairness, consistency, and robustness. A comprehensive list of these themes is provided in Section~\ref{sec:results}.

\begin{table}[t!]
\centering
\adjustbox{width=1.35\textwidth,center}{
\begin{tabular}{c c c c c c} 
\hline

ID & Tests Taken & Purposes of Tests & Roles & Gender & Native Languages \\
\hline
I1 & IELTS, PTE & Edu & Test-taker & Female & Mandarin, Cantonese \\
I2 & IELTS, PTE & Edu, Work, Immi & Test-taker & Male & Mandarin, Cantonese \\
I3 & TOEFL, IELTS, PTE & Edu, Immi & Test-taker & Male & Vietnamese \\
I4 & IELTS, PTE & Edu & Test-taker & Male & Mandarin \\
I5 & IELTS, PTE & Edu, Immi & Test-taker & Male & Spanish \\
I6 & IELTS, DET & Edu & Test-taker & Male & Mandarin \\
I7 & TOEFL, IELTS, PTE & Edu, Work, Immi & Test-taker, Teacher, Examiner & Female & Mandarin \\
I8 & IELTS, PTE & Edu, Work & Test-taker & Female & Mandarin \\
I9 & IELTS, PTE & Edu, Immi & Test-taker & Female & Mandarin, Cantonese \\
I10 & IELTS, PTE & Edu, Immi & Test-taker & Female & Mandarin \\
I11 & TOEFL, IELTS, PTE & Edu, Work, Immi & Test-taker, Teacher & Female & Korean \\
I12 & DET & Edu & Test-taker & Female & Mandarin \\
I13 & TOEFL & Edu & Test-taker & Female & Mandarin \\
I14 & TOEFL & Edu & Test-taker & Female & Mandarin \\
I15 & TOEFL, IELTS, DET & Edu & Test-taker & Male & Mandarin \\
I16 & IELTS, PTE & Edu, Immi & Test-taker & Female & Cantonese \\
       
 \hline
\end{tabular}
}
\caption{Information about interview participants. Education and Immigration in Purposes of Tests are denoted as Edu and Immi, respectively.}
\label{table:interviewees}
\end{table}

\subsection{Stage Two: Online Survey}
\label{sec:method:survey}

We conduct an online survey to gain a broader understanding of test-takers' experiences with English language tests (LTs). Based on the prior work~\cite{survey-guideline}, we design our online survey according to the following principles: setting clear objectives, selecting the appropriate design, formulating the survey questions, and gathering valid data. Notably, our survey responses are anonymous, ensuring that the information collected is non-identifiable. Additionally, human verification is also employed to prevent fraudulent responses, i.e., the responses that are not generated by English test-takers. The subsequent steps involved in this stage are listed as follows.

\noindent \textbf{Survey Design and Pilot Study.} 
We craft our online survey questions drawing from the findings obtained in the first stage interviews. This is done with the goal of achieving a broader comprehension of English test-takers' experiences. We utilize Qualtrics\footnote{\url{https://www.qualtrics.com/}}, an online survey platform, to distribute our survey. The survey comprises three question types: multiple choice, Likert scale, and rank order. We initially engage five participants for a pilot study, and their feedback subsequently informs the refinement of our survey's design and questions.

\noindent \textbf{Participant Recruitment.} 
In parallel with the first stage, we employ social media platforms for participant recruitment and also invite students to participate in our empirical study. In total, we collected 99 valid responses. Among these, 61 participants (61.6\%) had taken the IELTS, 29 (29.3\%) the TOEFL, 28 (28.3\%) the PTE, and nine participants (9.1\%) had experience with the DET. Our participants cover a broad demographic spectrum, providing diverse experiences. Specifically, the participants' native languages include Mandarin (67.7\%), Hindi (19.2\%), English (10.1\%), Cantonese (7.1\%), Spanish (2.0\%), and German (1.0\%). The motives for taking these tests span Education (89.9\%), Immigration (25.3\%), and Work (10.1\%).

\noindent \textbf{Online Survey.} 
Our online survey was open from November 1, 2022, to December 15, 2022. Throughout this period, we accumulated 165 responses. However, following the best practices from previous work~\cite{curran2016methods}, we manually filtered out responses that were deemed invalid due to factors such as duplicate IP addresses, unusually fast response times (completed within one minute), and inconsistent answers. As a result, our final dataset comprises 99 valid responses.

\noindent \textbf{Data Analysis.} 
We proceed to analyze the frequency of each option selected by the participants. Our objective is to identify potential relationships between the choices made across various questions and language tests. The findings from this analysis are presented in Section~\ref{sec:results}.

\section{Results}
\label{sec:results}
In this section, we delve into the findings of our empirical study. Initially, we shed light on English test-takers' concerns regarding AI-based and Human-based language tests (LTs). Subsequently, we scrutinize how the implementation of AI in LTs influences these test-takers.
\\
\\
\noindent \textbf{\ul{RQ1. What concerns do test-takers have regarding various types of language tests?}}
\label{rq1}

\noindent To address this research question, we offer our insights drawn from interviews and survey results, representing various perspectives on English language tests. Our findings indicate that test-takers possess distinct perceptions about LTs, particularly based on whether AI is utilized in the scoring process. However, opinions and concerns about LTs within the same category can vary significantly. Specifically, test-takers expressed concerns about the reliability of AI-based language tests (ALTs) and the fairness of Human-based language tests (HLTs). In this study, we classify PTE and DET as ALTs due to their dependence on AI for scoring. Conversely, other language tests, including TOEFL and IELTS, are referred to as HLTs, given their reliance on human effort for score determination.

\noindent\textbf{RQ1.1. Fairness and Consistency.}
In our interview stage (refer to Section~\ref{sec:method:interview}), 11 out of the 16 participants who took ALTs were of the opinion that the implementation of AI in automated scoring systems enhances the \textit{fairness} of LTs. This perception was corroborated by our online survey stage (refer to Section~\ref{sec:method:survey}), wherein 77.8\% of the ALT participants characterized these tests as ``unbiased.'' However, some interviewees did identify instances of perceived AI bias. For instance, participants I4, I5, and I10 (refer to Table~\ref{table:interviewees}) relayed the experiences of their acquaintances who consistently received low English test scores. The individuals referenced by I4 and I5 are high-pitched female test-takers from Asian countries, while those mentioned by I10 are male speakers with strong Spanish accents. These participants postulated that the AI models utilized in ALTs may struggle to recognize their specific pronunciations. Additionally, participants I7 and I8, who are Asian females, reported modifying their vocal pitch (deepening it) to secure higher scores on the PTE Speaking Test. I7 further speculated that the datasets used to train automated scoring systems may be imbalanced, causing a potential skew in fairness among diverse ethnic groups. This could imply preferential treatment for test-takers from certain ethnic groups over others.

In contrast, participants adopted a more critical view of the \textit{fairness} of HLTs during the interview phase. They voiced their apprehensions from various angles. For instance, I2 expressed concerns about the potential influence of human markers' moods on scoring, stating: \textit{``Even though the markers are highly trained and professional, their status during the marking may still potentially cause bias.''} Concurrently, I3 suggested that certain accents might not be favored by human markers, commenting: \textit{``From my own experience, if I keep my accent, the examiner will deduct a mark from my speaking test.''} These perceptions of HLT fairness were mirrored in our online survey. Specifically, only 31.1\% and 50\% of the participants deemed IELTS and TOEFL, respectively, as ``unbiased.''

During our interview stage, the majority of participants expressed a clear preference for the consistency provided by ALTs over HLTs. For instance, I8, who had experience with both test types, pointed out that while HLTs results varied significantly, ALTs outcomes remained stable within an acceptable range. This sentiment was echoed by I2, I6, I12, and I15, who also highlighted the consistency of AI-based LTs. Although a few participants maintained that HLTs were fair and consistent, the majority of those who had taken HLTs voiced their frustration regarding the fluctuation in their scores. For example, I5 and I14 characterized the results of HLTs as being \textit{``subjective and random''} and \textit{``unpredictable,''} respectively. Additionally, I10 and I16 shared their experiences about the inconsistent outcomes of HLTs across different English test centers. Our online survey results further corroborated the test-takers' perception that ALTs are more consistent than HLTs. Specifically, 32.15\% of survey participants who took ALTs perceived these language tests as consistent, in contrast to a mere 4.25\% of survey participants who took HLTs.

In summary, our findings suggest that participants perceive ALTs as fairer and more consistent in comparison to HLTs. Prior studies indicate that AI models employed in automated scoring systems are trained on a broad spectrum of responses sourced from English test-takers with varied educational backgrounds, nationalities, and accents~\cite{duolingo-paper, pte-white-paper}. Consequently, these AI models could potentially enhance overall fairness and consistency. Nevertheless, achieving balance among diverse groups within the training dataset remains a formidable challenge, which may contribute to an unavoidable bias in ALTs~\cite{ets-paper}. Conversely, professional human markers' assessments, subjected to their individual judgments and potential influence from external factors related to test-takers~\cite{ferman2021discriminating, ferman2022assessing, protivinsky2018gender}, can result in inevitable bias within HLTs.

\begin{tcolorbox}

\textbf{Finding RQ1.1.} 
Concerning fairness, our study's participant data generally suggests a perception of AI-based Language Tests (ALTs) as being more impartial and consistent than Human-based Language Tests. However, ALTs appear to demonstrate more bias in edge cases involving factors such as high-pitched voices, heavy accents, or specific ethnic groups.
\end{tcolorbox}

\noindent\textbf{RQ1.2. Cost, Availability, and Result Release Time.}
In our interview stage, a majority of the participants expressed satisfaction with the cost, availability, and result release time of ALTs over HLTs. Considering the cost, ALTs such as PTE and DET are priced at \$235 and \$59, respectively, while HLTs like TOEFL and IELTS require a minimum payment of \$255, varying based on location. Regarding availability, PTE stands out as it is offered daily, in contrast to the fortnightly schedule of TOEFL and IELTS. When considering result release time, ALTs usually take a maximum of two days, whereas HLTs can take up to a week. Participants also expressed their viewpoints on the availability and result release times of ALTs. For instance, I6 remarked that \textit{``booking an IELTS test and getting the result take very long, so I took DET to at least have a language test result on hand for university applications.''} Similarly, I7 emphasized that \textit{``the high availability and fast result release of PTE are very attractive for those who have limited time to get a satisfactory language test result.''} Supporting these views, our online survey indicated that 66.07\% of respondents were satisfied with the cost of ALTs, with only 24.36\% agreeing with the fee for HLTs. When assessing availability, 83.93\% favored ALTs, while 50.35\% were satisfied with HLTs. In terms of result release time, a significant 96.34\% of respondents were content with ALTs, compared to the lesser satisfaction level of 39.54\% for HLTs.

Our findings underscore the benefits of incorporating AI in LTs. From a cost perspective, the expense associated with developing, deploying, and maintaining AI models could potentially be more economical than the cost of training and employing professional human markers for scoring LTs. Moreover, ALTs offer a high degree of availability and produce results quickly, features that differentiate them from HLTs. Given that test-takers are sensitive to these factors, the majority of our study participants showed a preference for ALTs over HLTs.

\begin{tcolorbox}
\textbf{Finding RQ1.2.} AI-based Language Tests (ALTs), with their relatively lower costs, hold a significant appeal for English test-takers. Furthermore, in aspects such as availability and speed of result release, ALTs outperform Human-based Language Tests, making them a preferable choice.
\end{tcolorbox}

\noindent\textbf{RQ1.3. Robustness and Reliability.}
Prior research~\cite{test-taking-tricks} reveals that often adopt diverse strategies aimed at achieving higher scores. This practice can affect the reliability of tests and compromise the integrity of the evaluation systems. In the context of Language Tests (LTs), we observe a similar pattern where English test-takers employ various tactics to enhance their scores without necessarily demonstrating their genuine language proficiency.

During the interview stage, some participants disclosed employing various \textit{strategies} in AI-based Language Tests (ALTs) to secure higher scores. These strategies, popular among test-takers and language tutors, were mentioned by I1, I3, I4, I5, I6, I7, I8, I10, I12, and I16 as effective ways to improve their test scores. I3 noted, \textit{``applying tricks in PTE can largely improve the scores.''} I8 sincerely remarked, \textit{``I enrolled in a crash course simply to learn how to trick AI. If such tricks were not taught, I would directly drop the course.''} I4 and I10 expressed the effectiveness of these strategies, stating that AI is still \textit{``far from perfect''} and \textit{``unable to comprehend,''} respectively. Furthermore, our online survey showed that strategies for ALTs focused on enhancing speaking fluency (26.93\%), spoken discourse (15.28\%), written discourse (12.09\%), and speaking \& written content (15.28\%).

Our research reveals that participants taking HLTs also tend to utilize specific strategies to improve their scores; however, their views on the efficacy of these tactics are varied. During the interview stage, I4 suggested that human markers could potentially identify these strategies and consequently downgrade the test scores.\footnote{\url{https://www.scmp.com/comment/insight-opinion/united-states/article/2177403/how-english-testing-failing-chinese-students}} Despite this, a majority of the participants (I1, I3, I4, I6, I11, I14, and I16) taking HLTs are still inclined towards using these methods to enhance their English results. The data gathered from our online survey showed a higher propensity among ALT participants to employ strategies for better English scores. Notably, 55.36\% of ALT participants admitted using various strategies, compared to just 7\% of those taking HLTs.

We summarize the strategies utilized by participants in both ALTs and HLTs as follows:

\begin{itemize} [leftmargin=*]
    \item \textit{AI-based language tests:}
    \begin{itemize} [leftmargin=0.275cm]
        \item \textbf{Templates for essay writing (I1, I2, I3, I4, I6, I7, I8, I10, I12, I16).}
        Test-takers have gathered essay-writing templates that consist of pre-prepared content, including specific sentences, clauses, and logical connectives. These templates are memorized and deployed in the essay writing sections to enhance written discourse and content.
    
        \item \textbf{Templates for open-response speaking (I1, I3, I4, I5, I7, I8, I11, I12).}
        Test-takers prepare templates, incorporating key points, transitions, and reasons for open-ended speaking queries.\footnote{\url{https://webberz.in/blog/pte-describe-image-templates-to-achieve-high-score/}} During the exam, test-takers simply substitute the placeholders in these templates with words or sentences pertinent to the questions. By leveraging this pre-constructed content, test-takers can prevent pauses associated with formulating grammatically correct clauses or substantial content. The primary objective of this strategy is to enhance spoken discourse, speaking fluency, and speaking content.

        \item \textbf{Non-stop talking (I4, I5, I7, I11, I12, I16).}
        In an attempt to secure higher scores, test-takers reiterate their responses during speaking tests, a tactic designed to mislead AI models of automated scoring systems. This strategy aims to elevate the speaking content.
    
        \item \textbf{No self-correction (I2, I7, I8, I10, I12).}
        In speaking tests, when a slip of the tongue occurs, test-takers deliberately overlook these errors, avoiding self-correction to sustain the appearance of fluency. This strategy is geared towards improving speaking fluency.
    
        \item \textbf{Spitting keywords fluently (I7, I8, I9).}
        Test-takers consistently articulate keywords with fluidity while disregarding grammatical accuracy and logical consistency. This strategy is intended to enhance both speaking content and fluency.

        \item \textbf{Single-sentence response (I7, I8).}
        Despite the requirement to recite an entire passage during spoken language reproduction tasks, test-takers intentionally limit themselves to reading only a single sentence.\footnote{\url{https://www.youtube.com/watch?v=bRgc5cHKKp0}} This strategy aims to enhance speaking fluency.

        \item \textbf{Re-taking via disconnection (I6).}
        Test-takers manipulate the provision designed to address technical issues in DET tests.\footnote{\url{https://go.duolingo.com/securitywhitepaper}} Specifically, if they perceive their performance on prior questions as inadequate, they intentionally disrupt their internet connection to initiate a test restart, thereby increasing their chances of obtaining higher scores.
        
    \end{itemize}

    \item \textit{Human-based language tests:}
    \begin{itemize} [leftmargin=0.275cm]
        \item \textbf{Templates for essay writing (I2, I4, I7, I8, I9, I11, I13, I14).}
        This strategy is similar to ALTs.

        \item \textbf{Hitting word count (I6, I7, I11, I13, I14).}
        Test-takers endeavor to produce as much text as feasible while adhering to the maximum word count in an effort to secure higher scores. The goal of this strategy is to enhance the assessment of content in writing.
        
        \item \textbf{Templates for open-response speaking (I1, I3, I4, I5, I7, I8, I11, I12).}
        This strategy is similar to ALTs.
    \end{itemize}
\end{itemize}

Participants sourced their strategies from various platforms, including language schools (I1, I8, I10), fellow test-takers (I5 and I13), and online resources (I6, I9, I12, I14, and I16). Additionally, I7, an English tutor, routinely took English tests to verify the effectiveness of these strategies. In summary, our findings reveal the widespread adoption of multiple strategies in both ALTs and HLTs to enhance test scores. Templates are a prevalent approach for both ALTs and HLTs. Nonetheless, ALTs have unique strategies, such as continuous speaking and avoidance of self-correction. These tactics target the vulnerabilities of AI models in automated scoring systems, specifically their robustness and reliability, thereby aiding test-takers in achieving superior AI-based language test scores.

\begin{tcolorbox}
\textbf{Finding RQ1.3.}
Test-takers of both AI-based language tests (ALTs) and Human-based language tests implement various strategies to optimize their performance, particularly in the speaking and writing sections. Furthermore, test-takers of ALTs specifically endeavor to mislead the AI models used in automated scoring systems, intending to secure higher scores.
\end{tcolorbox}

\noindent\textbf{RQ1.4. Transparency and Explainability.}
Participants discussed the \textit{transparency} and \textit{explainability} of language tests (LTs), focusing on marking metrics, credibility, and the clarity of explanations. Despite the availability of published details on marking metrics from the test organizers, some participants (I4 and I11) expressed a lack of understanding of these metrics. Most interviewees obtained information about the marking metrics from third-party sources, such as language schools (I1, I8, I12, I13, and I14), online resources (I2, I3, I6, I9, I10, I11, I12, I13, I14, and I16), friends (I4 and I5), and books (I5). The trustworthiness of the LTs was also questioned by participants, even in the face of assurances from test organizers.\footnote{\url{https://www.ielts.org/-/media/publications/quality-and-fairness/quality-and-fairness-2015-uk.ashx}}\footnote{\url{https://www.ets.org/toefl/research/reliability-validity.html}}\footnote{\url{https://www.pearson.com/content/dam/one-dot-com/one-dot-com/global/Files/efficacy-and-research/reports/PTE-Academic-Assessment-Efficacy-Report-2019.pdf}}\footnote{\url{https://blog.duolingo.com/fairness/}} In the case of ALTs, some participants (I4 and I10) were unsure about the workings of automated scoring systems, despite the availability of this information in the public domain through academic papers~\cite{duolingo-paper} or online media.\footnote{\url{https://www.pearsonpte.com/scoring/automated-scoring}} Test-takers of both ALTs and HLTs expressed a desire for their results to include more explanatory details to help them improve their language proficiency. However, the feedback they received was typically general, showing only broad score ranges without personalized input. This lack of detailed feedback led some participants (I3, I5, I11, and I12) to dismiss the information as ``useless,'' while others (I5 and I6) expressed disappointment about the scanty explanation provided in their test results.

Furthermore, some participants expressed confusion regarding the scoring methods used in AI-based language tests (ALTs). For example, PTE and DET employ unique scoring methodologies instead of a simple summation of all section scores. In PTE, each question evaluates multiple communicative skills simultaneously, with the performance on each question contributing to the final scores of different communicative skill sections. Additionally, the overall score is not a mere average of all communicative skill section scores. Participant I1 reported confusion when her overall score was lower than the average of all section scores. Similar experiences have also been documented online, with individuals sharing screenshots showing an overall score lower than any individual communicative skill section score.\footnote{\url{https://www.xiaohongshu.com/explore/6380e1460000000018013afb}} On the other hand, DET uses an adaptive testing method~\cite{duolingo-adaptive}, where the final scores of the test-takers are dependent not only on the correctness of their answers but also on the difficulty level of each question. The difficulty level adapts based on the test-taker's performance on previous questions. This adaptive testing approach, as participants I6 and I12 observed, is perplexing and can induce anxiety and uncertainty during the tests due to its fluid nature.

In summary, our findings reveal a prevalent lack of transparency and explainability in both ALTs and HLTs. Despite the availability of detailed information on the websites or guidelines of language test organizers, test-takers often find it challenging to understand these materials. Furthermore, there is a clear expectation among test-takers for well-explained test results. They believe such comprehensive feedback is essential for their understanding and improvement in language proficiency.

\begin{tcolorbox}
\textbf{Finding RQ1.4.}
Test-takers anticipate both transparency and explainability in AI-based and Human-based language tests. However, the current level of information provided often falls short of these expectations.
\end{tcolorbox}

\noindent\textbf{RQ1.5. Contestability and Accountability.}
In both AI-based and Human-based language tests, official processes exist for test-takers to challenge their results. However, the experiences of our participants varied considerably when submitting an appeal. Some participants (I3, I6, I9, and I14) expressed reluctance to appeal due to the potential costs, the likelihood of unchanged outcomes, or even the fear of point reduction. I3 mentioned, \textit{``the score was remarkably lower than I expected, but according to the rules of IELTS, even if the score were fixed, it would be increased by 0.5 at most, which was still not ideal to me, so I just didn't do it.''} Both I5 and I6 reported that their appeals did not lead to any changes in their scores. Overall, these findings suggest that test-takers harbor reservations about the fairness and contestability of language tests. The potential financial burden and fear of point reduction discourage them from contesting the results, thereby reducing the likelihood of test organizers being held accountable.

\begin{tcolorbox}
\textbf{Finding RQ1.5.}
Due to factors such as substantial costs, perceived low probability of success, and potential score reductions, test-takers are often disinclined to challenge their results or hold language test organizers accountable.
\end{tcolorbox}

\noindent\textbf{RQ1.6. Other Concerns.}
During our interview stage, participants expressed several additional concerns, particularly related to the usage of the collected information. For instance, interviewee I1 highlighted that PTE used to include a ``first-time test taker'' label on the result report, this label even appears on the official sample score report.\footnote{\url{https://www.pearsonpte.com/ctf-images/yqwtwibiobs4/5Tkz5xNp0H67FtzmLbP8yX/5d9493a87e90e1cbdbe63117b479a564/score\_report.png}} The participants voiced confusion about the purpose of this label and whether it led to differential treatment of test-takers. Further concerns were raised about the potential for language test providers to leverage algorithms to induce a ``near-miss'' effect~\cite{near-miss}, thereby encouraging test-takers to retake the tests more frequently.

Additional participants, I2 and I12, expressed concerns about the profit-oriented nature of test providers. They shared the prevalent belief among test-takers and tutoring organizations that achieving comparable scores on the PTE or DET is apparently easier compared to the IELTS and TOEFL. These interviewees were concerned that this might be a marketing strategy designed to sway test-takers toward these tests. Interviewee I8 recounted a similar experience, describing how a conversation with a fellow passenger on a flight led them to switch from the IELTS to the PTE. They stated: \textit{``On the airplane, I talked with the girl next to me about the language tests we took. I felt very frustrated about taking IELTS. She told me that if I had taken PTE, I would not feel frustrated. I later took PTE and felt she was right. It is too easy to get high scores in PTE.''}
Moreover, test-takers highlighted the non-interactive aspect of the speaking component in ALTs and the TOEFL. While certain participants (I3 and I4) favored these non-interactive speaking tests due to shyness, others maintained that \textit{``language is about communicating with other humans''} (I11). Therefore, they felt that speaking to a computer without engaging in authentic dialogue was not representative of real-world language use (I2, I11, and I16).

\begin{tcolorbox}
\textbf{Finding RQ1.6.}
Participants in both AI-based and Human-based language tests expressed additional concerns, including the profit-driven motives of test providers, the validity of test results, and the lack of interactivity during the test.
\end{tcolorbox}

\noindent \textbf{\ul{RQ2. What is the impact of using AI in language tests on test-takers?}}

\noindent As an increasing number of institutions, organizations, and governments accept language tests (LTs), test-takers have more freedom to select their preferred test. In this research question, we aim to explore how the incorporation of AI impacts test-takers and how these individuals enhance their language proficiency during their test preparation period.

\begin{figure}[t!]
  \centering
\adjustbox{center}{
  \includegraphics[width=1.15\textwidth]{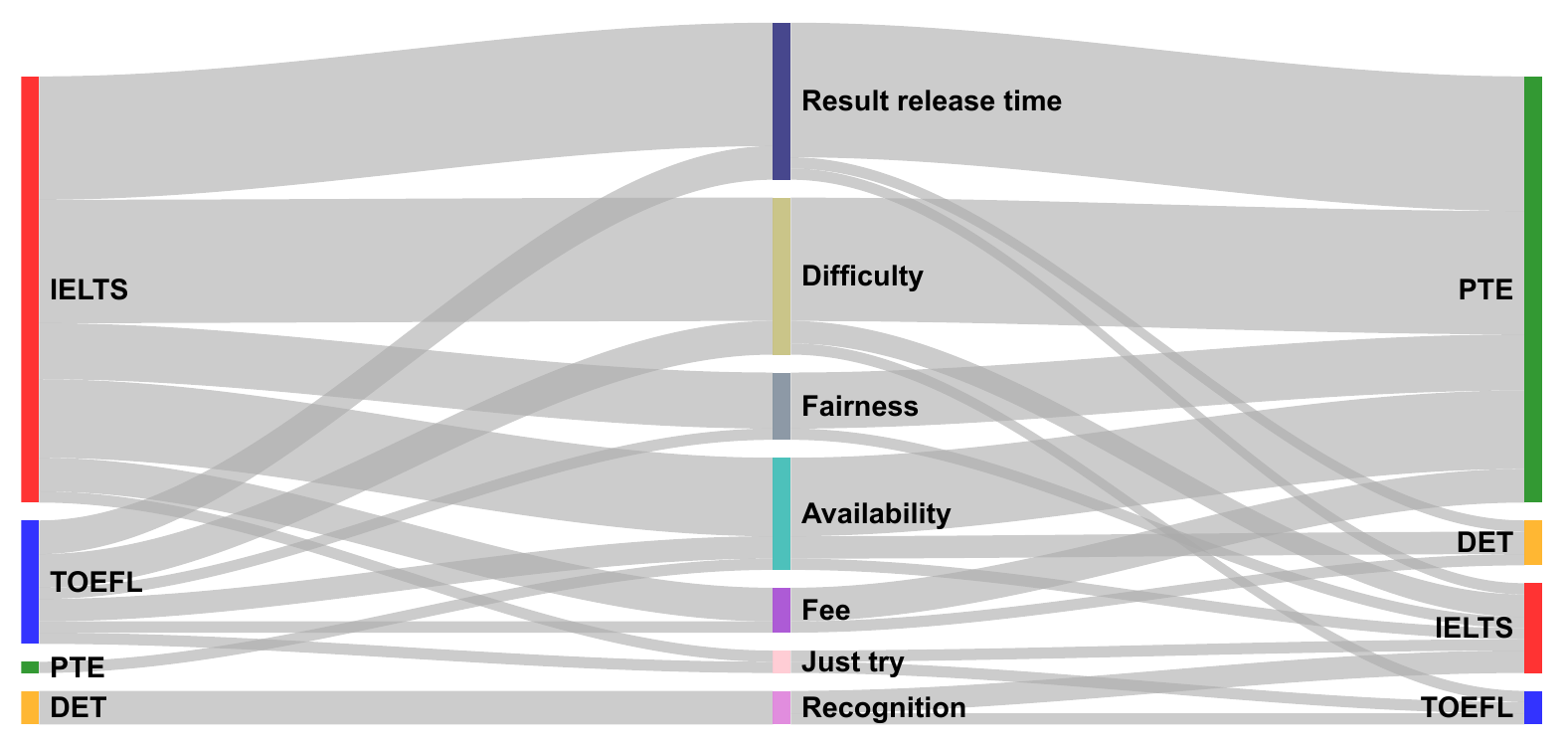}
  }
  \caption{Sankey diagram on the change of test choices based on the results of online survey. The left column shows the departure tests, and the right column shows the destination tests. The middle column shows their reasons of changing the test choices.}
  \label{fig:test-change}
\end{figure}

\noindent\textbf{RQ2.1. Choices of Tests.}
From our online survey, we observed that 24 out of 99 participants indicated having taken multiple LTs. Figure~\ref{fig:test-change} illustrates the reasons and dynamics behind participants' shifting language test choices, as derived from our survey data. We found that a substantial number of participants who switched tests transitioned from IELTS to PTE (17 participants). Factors such as result release time, test difficulty, and test availability significantly influenced this change. Furthermore, our survey revealed a preference for ALTs, such as PTE, among participants intending to immigrate, with only three out of the 14 potential immigrants choosing HLTs.

ALTs can partially alleviate the financial and time constraints for test-takers, but this could potentially spark more intense competition among them. Given that spots in highly respected universities are often limited, these institutions typically operate on a merit-based, first-come, first-served basis. The advantages of ALTs could exacerbate this competitive atmosphere. As an illustration, one of our interviewees, I7, an English tutor, highlighted this predicament, noting that students were compelled to \textit{``test early, frequently, and repeatedly.''}

\begin{tcolorbox}
\textbf{Finding RQ2.1.}
Test-takers are increasingly opting for AI-based language tests, prompted by factors, such as faster result release times, perceived difficulty, and broad availability.
\end{tcolorbox}

\begin{figure}[t!]
  \centering
\adjustbox{center}{
  \includegraphics[width=1.2\textwidth]{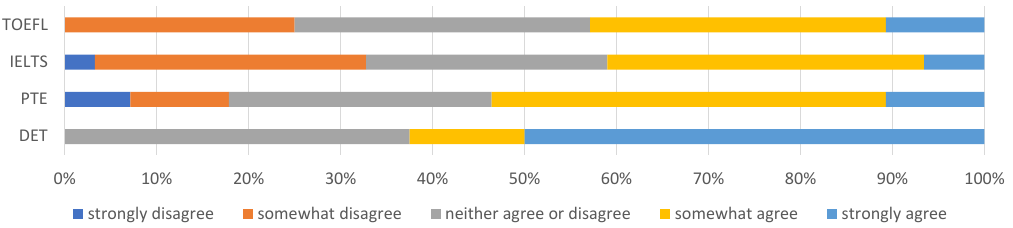}
  }
  \caption{The participants' agreement on the usefulness of test reports for language improvement}
  \label{fig:report-usefulness}
\end{figure}

\noindent\textbf{RQ2.2. Language Improvement.}
Test-takers typically use their language test scores as indicators to improve their language proficiency. While they expect the test results to guide their language proficiency improvement, many express dissatisfaction with the quality of feedback provided in the test results. Figure~\ref{fig:report-usefulness} displays the participants' consensus on the usefulness of the language test results for their language enhancement. Participants in ALTs, such as PTE and DET, generally provide positive feedback about the benefits of the test results. In contrast, participants from HLTs, including TOEFL and IELTS, present a more neutral viewpoint.

\begin{figure}[t!]
  \centering
\adjustbox{center}{
  \includegraphics[width=1.2\textwidth]{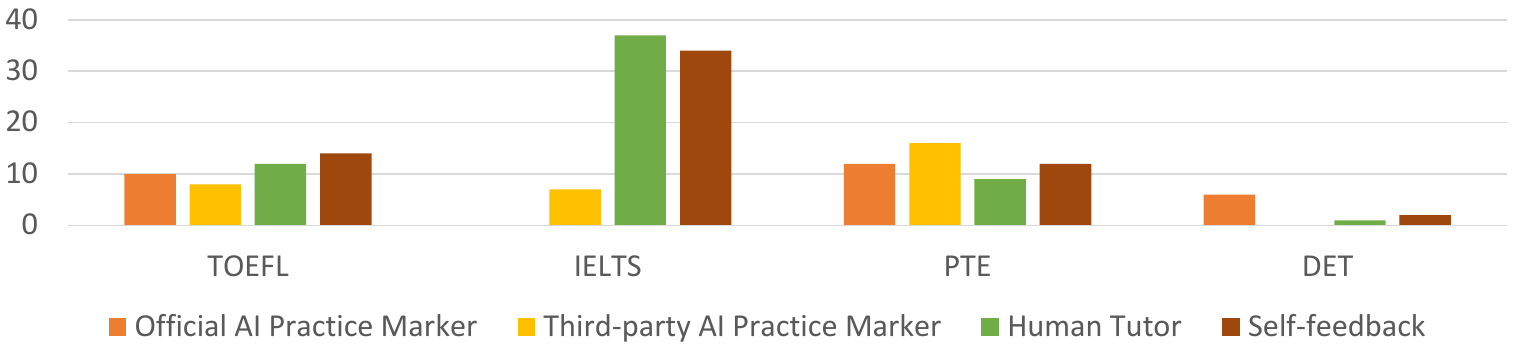}
  }
  \caption{The participants' choices of practice methods}
  \label{fig:practice-methods}
\end{figure}

Figure~\ref{fig:practice-methods} presents practice methods, including an official AI practice marker, a third-party AI practice marker, a human tutor, and self-feedback, that participants utilize in preparation for LTs in our online survey stage. Note that IELTS has no official AI practice marker. The figure reveals that participants preparing for ALTs and HLTs significantly rely on the human tutor and self-feedback. In particular, when rating on a 5-point Likert scale (where 4 or 5 indicates strong agreement), 80.74\% and 88.89\% of participants in ALTs and HLTs believe that the human tutor significantly enhances their test scores, respectively. Specifically, the human tutor scored the highest (3.95 for ALTs and 4.11 for HLTs), surpassing other methods, such as the official AI practice marker, third-party AI practice marker, and self-feedback.

During our interview stage, participants (I2, I7, I8, and I13) shared that employing the human tutor or acquiring the official AI practice marker was financially burdensome, leading them to opt for the third-party AI practice marker—a preference also illustrated in Figure~\ref{fig:practice-methods}. However, the third-party AI practice marker's extensive usage could negatively affect language learning effectiveness due to its misalignment with actual language proficiency (as noted by I1 and I2). Furthermore, I7 pointed out that if English test-takers implement the strategies discussed in RQ1 to enhance their scores—which may involve unnatural usage—this could potentially detriment their actual language proficiency.

\begin{tcolorbox}
\textbf{Finding RQ2.2.}
Based on findings from participants and an online survey, a human tutor and an official AI practice marker proved beneficial for enhancing test-takers' language proficiency. However, their high cost leads test-takers to opt for a third-party AI marker, a choice that may yield less benefit or even produce adverse effects.
\end{tcolorbox}

\section{Discussion}

In this section, we discuss the primary implications derived from our empirical findings related to the use of AI in language testing.

Our empirical study reveals that AI-based Language Tests (ALTs) have gained popularity among test-takers for various reasons, including consistency, rapid result release times, cost-effectiveness, and widespread availability. Furthermore, educational establishments, professional accreditation bodies, and governments are increasingly accepting ALT results. Specifically, as of February 1, 2023, Canada joined countries such as Australia, New Zealand, and the United Kingdom in accepting PTE scores for immigration purposes,\footnote{\url{https://www.canada.ca/en/immigration-refugees-citizenship/corporate/publications-manuals/operational-bulletins-manuals/updates/2023-designated-language-testing-organization.html}} signaling the growing acceptance of ALTs. However, Human-based Language Tests (HLTs) still hold the edge over ALTs in terms of robustness and reliability. Despite considerable research into the trustworthiness of AI models over the past decade, many issues remain unresolved. Consequently, automated scoring systems' AI models may inherit these problems in language testing, leading to societal implications. For instance, imbalanced training data might exacerbate biases towards individuals from specific groups, undermining societal diversity, equity, and inclusion. Furthermore, test-taking strategies designed to maximize scores could compromise the validity of language tests as a merit-based assessment tool. As AI models of ALTs are often operated on a large scale, their impact on the community may be more profound and systematic compared to HLTs.

The trustworthiness of language tests is crucial; however, limited transparency restricts access to such information. Even though AI-based language test organizers disclose papers or audit reports about their tests, their models and training data largely remain undisclosed, possibly due to commercial interests. By releasing datasheets~\cite{gebru2021datasheets} or model cards~\cite{mitchell2019model} without necessarily making them public, they could help inform relevant stakeholders about their trustworthiness. Furthermore, based on participant feedback, official AI practice markers have proven beneficial for test preparation and familiarization with language tests alongside traditional human tutoring. These AI practice markers also aid test-takers in comprehending the trustworthiness of AI models in automated scoring systems. However, their high cost deters test-takers from utilizing the official AI practice markers, leading them to resort to third-party AI practice markers, which could potentially impede their language learning progress. Our finding reveals that AI-based and Human-based language test organizers have a significant distance to cover in maintaining transparency and fostering trust among test-takers.

With the recent advancements in natural language understanding, more sophisticated technologies, specifically large language models (LLMs), have been developed for various applications, including practice interviews,\footnote{\url{https://grow.google/certificates/interview-warmup/}} AI assistants,\footnote{\url{https://callannie.ai/}} and language learning assistants.\footnote{\url{https://www.speak.com/}} Given the close correlation between language testing and language models, there is considerable potential for these technologies to be integrated into ALTs. Since the reliability of language tests carries significant societal implications, LLMs are not exempt from issues such as bias and a lack of explainability. Therefore, these LLMs must undergo comprehensive evaluations before deployment in practice.

\section{Threats to Validity}

While we adhered to established practices in designing our empirical study, the clarity of our questions might not have been optimal, potentially skewing the analysis of our dataset. In addition, we had no control over the authenticity of participant responses. To mitigate this problem, we manually filtered out invalid responses; however, some might remain. To facilitate a wide range of participant experiences and backgrounds, we employed various social media platforms to recruit English test-takers for our study. Nevertheless, the test-taker composition was imbalanced across various spectrums, leading to inherent biases in our collected dataset.

Our empirical study focused on the four major English language tests, including TOEFL, IELTS, PTE, and DET, as of February 2023. Therefore, our findings may not be generalizable beyond these specific language tests and timeframes. Moreover, our dataset, collected exclusively from English test-takers, may not reflect the perspectives of other stakeholders in the language testing domain.

\section{Conclusion}
Educational institutions, professional accreditation bodies, and governments frequently utilize language tests to gauge candidates' proficiency, informing merit-based selections or decisions. Recent advancements in AI technology have prompted test providers to leverage AI capabilities in developing automated scoring systems. As the trustworthiness of AI increasingly comes under scrutiny, the implications of AI usage in language tests require thorough exploration to comprehend their impacts on individuals and the larger community. To our knowledge, this research represents the first empirical study exploring these implications from the perspective of test-takers.

Our findings suggest that AI-based language tests (ALTs) are generally perceived as fairer and more consistent than Human-based language tests (HLTs), although they may exhibit more bias in certain scenarios, such as with high-pitched voices or strong accents. The allure of ALTs lies in their cost-effectiveness, superior accessibility, and speedy result release times, making them the preferred choice for many test-takers. In terms of reliability and robustness, test-takers of both ALTs and HLTs utilize a variety of strategies to attain higher scores, with AI-based language test-takers more likely to employ these strategies to mislead AI models within automated scoring systems. Despite a clear desire for transparency and explainability among test-takers, such information is often insufficiently provided. While human tutors and official AI practice markers can aid in improving test-takers' language proficiency, their high costs often lead test-takers to resort to third-party AI markers, which could potentially hinder their learning progress. In an era where AI is disrupting traditional patterns of language testing and learning, we present empirical evidence of relevant issues, raising awareness of their potential societal implications. We also advocate for stakeholders to fully comprehend these implications prior to embracing the use of AI.

\end{document}